\documentclass[preprint,showpacs,preprintnumbers,amsmath,amssymb,nofootinbib]{revtex4}
\usepackage{booktabs}
\usepackage{mathrsfs}
\usepackage{epsfig}
\usepackage{graphicx}
\usepackage{dcolumn}
\usepackage{bm}
\usepackage{amsmath}
\usepackage{slashed}
\usepackage{multirow}
\usepackage{color}

\let\jnfont=\rm
\def\NPB#1,{{\jnfont Nucl.\ Phys.\ B }{\bf #1},}
\def\PLB#1,{{\jnfont Phys.\ Lett.\ B }{\bf #1},}
\def\EPJC#1,{{\jnfont Eur.\ Phys.\ Jour.\ C }{\bf #1},}
\def\PRD#1,{{\jnfont Phys.\ Rev.\ D }{\bf #1},}
\def\PRL#1,{{\jnfont Phys.\ Rev.\ Lett.\ }{\bf #1},}
\def\MPLA#1,{{\jnfont Mod.\ Phys.\ Lett.\ A }{\bf #1},}
\def\JPG#1,{{\jnfont J.\ Phys.\ G}{\bf #1},}
\def\CTP#1,{{\jnfont Commun.\ Theor.\ Phys.\ }{\bf #1},}
\def\ZPC#1,{{\jnfont Z.\ Phys.\ C }{\bf #1},}
\def\JHEP#1,{{\jnfont JHEP \ }{\bf #1},}
\def\Rv{\not{\hbox{\kern-1pt $R$}}}
\def\p{\not{\hbox{\kern-3pt $p$}}}

\newcommand{\bea}{\begin{eqnarray}}
\newcommand{\eea}{\end{eqnarray}}

\newcommand{\bcen}{\begin{center}}
\newcommand{\ecen}{\end{center}}

\newcommand{\beq}{\begin{eqnarray}}
\newcommand{\eeq}{\end{eqnarray}}

\def\t1{\tilde{t_1}}
\def\tst{\tilde t}
\begin{document}

\preprint{
\begin{minipage}[b]{0.75\linewidth}
\begin{flushright}
IPMU16-0130 \\
CTPU-16-26
 \end{flushright}
\end{minipage}
}

\title{Top-squark in natural SUSY under current LHC run-2 data}
\author{ Chengcheng Han$^{1}$}
\author{ Jie Ren$^{2}$}
\author{ Lei Wu$^{3,4}$}
\author{ Jin Min Yang$^{5}$}
\author{ Mengchao Zhang$^{6}$
\vspace*{.5cm}}

\affiliation{
$^1$ Kavli IPMU (WPI), UTIAS, University of Tokyo, Kashiwa, 277-8583, Japan \\
$^2$ Computer Network Information Center, Chinese Academy of Sciences, Beijing 100190, China\\
$^3$ Department of Physics and Institute of Theoretical Physics, Nanjing Normal University, Nanjing, Jiangsu 210023, China\\
$^4$ ARC Centre of Excellence for Particle Physics at the Terascale, School of Physics, The University of Sydney, NSW 2006, Australia\\
$^5$ Key Laboratory of Theoretical Physics, Institute of Theoretical Physics, Chinese Academy of Sciences, Beijing 100190, China\\
$^6$ Center for Theoretical Physics and Universe, Institute for Basic Science (IBS),
Daejeon 34051, Korea.
}%


\begin{abstract}

We utilize the recent LHC-13 TeV data to study the lower mass bound on top-squark (stop) in natural supersymmetry.
We recast the LHC sparticle inclusive search of $(\ge 1){\rm jets} + \slashed E_T$ with $\alpha_T$ variable,
the direct stop pair search (1-lepton channel and all-hadronic channel) and the monojet analyses.
We find that
these searches are complementary depending on stop and higgsino masses:
for a heavy stop the all-hadronic stop pair search provides the strongest bound,
for an intermediate stop the inclusive SUSY analysis with $\alpha_T$ variable is most efficient,
while for a compressed stop-higgsino scenario the monojet search plays the key role.
Finally, the lower mass bound on a stop is: (i) 320 GeV for compressed stop-higgsino scenario
(mass splitting less than 20 GeV); (ii) 765 (860) GeV for higgsinos lighter than 300 (100) GeV.
\end{abstract}

\pacs{12.60.Jv, 14.80.Ly}
\maketitle

\section{INTRODUCTION}
The discovery of the Higgs boson is a great triumph for the Standard Model (SM). However, the SM Higgs mass is
quadratically sensitive to the cutoff scale $\Lambda$ (usually taken as GUT or Planck scale) via radiative corrections
because of the lack of symmetry protection. This renders the SM with $m_h \sim 125~{\rm GeV} \ll \Lambda$ rather
unnatural. A well known theory of solving such a naturalness problem is supersymmetry.

Among various supersymmetric models, natural supersymmetry (NSUSY) is a well motivated framework
\cite{nsusy-1,nsusy-2,nsusy-3},
which consists of a small set of sparticles that closely
relate to the naturalness, such as higgsinos, stop and gluino. This can be understood by the minimization
of the Higgs potential \cite{mz}
\begin{eqnarray}
\frac{M^2_{Z}}{2}&=&\frac{(m^2_{H_d}+\Sigma_{d})-(m^2_{H_u}+
\Sigma_{u})\tan^{2}\beta}{\tan^{2}\beta-1}-\mu^{2} \nonumber \\
&\simeq& -(m^2_{H_u}+
\Sigma_{u})-\mu^{2} ,
\label{minimization}
\end{eqnarray}
where $\mu$ is the higgsino mass parameter in the superpotential and contributes to $M_Z$ at tree level,
$\tan\beta \equiv v_u/v_d \gg 1$ is assumed in the last approximate equality, $m^2_{H_d}$ and $m^2_{H_u}$ denote
the soft SUSY breaking masses of the Higgs fields at weak scale, and $\Sigma_{u}$ and $\Sigma_{d}$ arise from
the radiative corrections to the Higgs potential. Due to the large top Yukawa couplings, $\Sigma_{u}$ is dominated
by the stop at 1-loop level, while the gluino contributes to $\Sigma_{u}$ via the corrections to the stop mass.
Other contributions from the first two generation squarks and sleptons to $M_Z$ are negligible small.
Therefore, the requirement of getting the correct value of $M_Z$ without fine-tuning will give upper bounds on
the masses of higgsinos, stops and gluino \cite{bg,baer-upper-bound}. In the past few years, many works have been devoted to the searches for the stop at the LHC in NSUSY \cite{nsusy-4,nsusy-4.5,nsusy-5,nsusy-6,nsusy-7,nsusy-8,wu-0,nsusy-10,wu-2,warsaw,nsusy-11,wu-1,wu-3,kim-1,nsusy-12,wu-4}.

With the recent $\sim 15 fb^{-1}$ dataset at the LHC run-2, the stop and gluino masses are
respectively excluded up to $\sim$ 1 TeV \cite{run2-stop} and 1.8 TeV \cite{run2-gluino}, while the electroweakinos below
$0.4-1$ TeV can also be covered for different decay channels \cite{run2-electroweakino}. But these limits
are obtained in the simplified models \cite{run2-1,run2-2,run2-3} and sensitively depend on the assumptions of the nature of the lighest
supersymmetric partner (LSP), the branching ratios of heavier sparticles and the mass splitting between
heavier sparticles and the LSP. Therefore, it is necessary to examine the current LHC run-2 coverage of NSUSY
and assess the fine-tuning extent. In this work, we utilize the recent results of the LHC run-2 inclusive
sparticle searches and direct stop pair searches to constrain the stop mass in NSUSY. We compare their sensitivities
and find that they are complementary in probing NSUSY. We will also evaluate the electroweak fine-tuning measure
in the allowed parameter space of NSUSY and comment on the prospect for covering the low fine-tuning parameter
space of NSUSY at HL-LHC.

\section{Constraints on Stop in NSUSY}\label{section2}
In MSSM, the stop mass matrix in the weak-basis $(\tilde{t}_L, \tilde{t}_R)$ is given by
\begin{equation}
M_{\tilde t}^2 =\left(\begin{array}{cc}
m_{{\tilde t}_L}^2& m_tX_t^\dag\\
 m_tX_t& m_{{\tilde t}_R}^2 \end{array} \right) \ ,
 \label{stop_weak}
\end{equation}
with
\begin{eqnarray}
m_{{\tilde t}_L}^2 &=& m_{\tilde{Q}_{3L}}^2+m_t^2+m_Z^2(\frac{1}{2}
-\frac{2}{3}\sin^2\theta_W)\cos2\beta \ , \\
m_{{\tilde t}_R}^2 &=& m_{\tilde{U}_{3R}}^2+m_t^2 +\frac{2}{3}m_Z^2 \sin^2\theta_W\cos2\beta \ , \\
X_t&=& A_t-\mu\cot\beta \ .
\end{eqnarray}
Here $m_{\tilde{Q}_{3L}}^2$ and $m_{\tilde{U}_{3R}}^2$ are the soft-breaking mass parameters
for the third generation left-handed squark doublet $\tilde{Q}_{3L}$ and the right-handed stop $\tilde{U}_{3R}$, respectively. $A_t$ is the stop soft-breaking trilinear parameter. The weak eigenstates $\tilde t_{L,R}$ can be rotated to the mass eigenstates $\tilde t_{1,2}$ by a unitary transformation,
\begin{eqnarray}
\left (\begin{array}{c} \tilde t_1 \\ \tilde t_2 \end{array} \right )
= \left ( \begin{array}{cc}\cos\theta_{\tilde{t}} &\sin\theta_{\tilde{t}} \\
-\sin\theta_{\tilde{t}} &\cos\theta_{\tilde{t}} \end{array} \right )
\left (\begin{array}{c} \tilde t_L \\ \tilde t_R \end{array} \right).
\end{eqnarray}
After diagonalizing the mass matrix Eq.~(\ref{stop_weak}), we can have the stop masses $m_{\tilde t_{1,2}}$ and the mixing angle $\theta_{\tilde{t}}$ ($-\pi/2 \leq \theta_{\tilde{t}} \leq \pi/2$),
\begin{eqnarray}
m_{\tilde t_{1,2}}&=&\frac{1}{2}\left[ m_{{\tilde t}_L}^2+m_{{\tilde t}_R}^2
\mp\sqrt{\left(m_{{\tilde t}_L}^2-m_{{\tilde t}_R}^2\right)^2 +4m_t^2X_t^2}\right] ,\\
\tan2\theta_{\tilde{t}} &=& \frac{2m_tX_t}{m_{{\tilde t}_L}^2-m_{{\tilde t}_R}^2} \ .
\end{eqnarray}
The decays of stop are determined by the interactions between stop and neutralinos/charginos, which are
given by
\begin{eqnarray}
    {\cal L}_{\tilde{t}_1\bar{b}\tilde{\chi}^+_i} &=& \tilde{t}_1
    \bar{b} ( f^{C}_L P_L + f^{C}_R P_R ) \tilde{\chi}^+_i +h.c.~, \\
    {\cal L}_{\tilde{t}_1\bar{t}\tilde{\chi}^0_i} &=& \tilde{t}_1
    \bar{t} ( f^{N}_L P_L + f^{N}_R P_R ) \tilde{\chi}^0_i + h.c.~,
    \label{vertex}
\end{eqnarray}
where $P_{L/R}=(1\mp\gamma_5)/2$ and
\begin{eqnarray}
    &&f^N_L =
    -\left[ \frac{ g_2 }{\sqrt{2}}N_{i 2}
        + \frac{ g_1}{3\sqrt{2}} N_{i 1}
    \right] \cos\theta_{\tilde{t}} -y_t N_{i 4} \sin\theta_{\tilde{t}}~~~~
\label{left-handed} \\
    &&f^N_R = \frac{2\sqrt{2}}{3} g_1
    N^*_{i 1} \sin\theta_{\tilde{t}}
    - y_t N^*_{i 4} \cos\theta_{\tilde{t}},\\
    &&    f^C_L = y_b U^*_{i 2} \cos\theta_{\tilde{t}},\\
    && f^C_R = - g_2 V_{i 1} \cos\theta_{\tilde{t}}
    + y_t V_{i 2}\sin\theta_{\tilde{t}} ,
\label{right-handed}
\end{eqnarray}
with $y_t=\sqrt{2}m_t/(v\sin\beta)$ and $y_b=\sqrt{2}m_b/(v\cos\beta)$ being the Yukawa couplings of top and bottom quarks. The mixing matrices of neutralinos $N_{ij}$ and charginos $U_{ij}$, $V_{ij}$ are defined in \cite{mssm-feynrules}. In NSUSY, $M_{1,2}\gg \mu$, one has $V_{11}, U_{11}, N_{11,12,21,22} \sim 0$, $V_{12} \sim \mathop{\rm sgn}(\mu)$, $U_{12} \sim 1$ and $N_{13,14,23}=-N_{24} \sim 1/\sqrt{2}$. So, $\tilde{\chi}^{\pm}_{1}$ and $\tilde{\chi}^0_{1,2}$ are higgsino-like and nearly degenerate \footnote{The detection of such light higgsinos through monojet(-like) may be challenging at the LHC \cite{giudice-higgisno,baer-higgsino,wu-higgsino,han-higgsino,park-higgsino}.}. The left-handed stop will mainly decay to $t \tilde{\chi}^{0}_{1,2}$ when the phase space is accessible and $\tan\beta$ is small. While the couplings of the right-handed stop with
$\tilde{\chi}^0_{1,2}$ and $\tilde{\chi}^\pm_{1}$ are proportional to $y_t$, and the branching ratios of $\tilde{t}_{1} \to t \tilde{\chi}^{0}_{1,2}$ and $\tilde{t}_1 \to b \tilde{\chi}^+_1$ are about 25\% and 50\%, respectively.

To address the lower mass limit of stop in NSUSY, we can focus on a right-handed stop. This is because that the
left-handed stop is linked with the left-handed sbottom by the $SU(2)$ symmetry. Then, the left-handed sbottom
decay channel $\tilde{b}_1 \to t \tilde{\chi}^-_1$ can mimics the left-handed stop signals $\tilde{t}_{1} \to t \tilde{\chi}^{0}_{1,2}$ since $\tilde{\chi}^0_{1,2}$ and $\tilde{\chi}^+_{1}$ are higgsino-like and degenerate in NSUSY.
This enhances the LHC limit on a left-handed stop, which is stronger than the limit on
a right-handed stop \cite{wu-2,kim-1}.

Now we examine the constraints on the NSUSY scenario that consists of a right-handed stop and higgsinos.
We scan the parameter space in the ranges
\begin{eqnarray}
100~{\rm GeV} &\le& \mu \le 600~{\rm GeV}, \quad 100~{\rm GeV} \le m_{\tilde{Q}_{3L},\tilde{U}_{3R}}\le 2.5~{\rm TeV}, \nonumber \\ 1~{\rm TeV} &\le& A_t \le 3~{\rm TeV}, \quad 5 \le \tan\beta \le 50.
\end{eqnarray}
The lower limit on the higgsino mass is motivated by the LEP searches for electroweakinos. We require the stop mixing
angle $|\sin\theta_{\tilde{t}}|^2>0.5$ to obtain a right-handed stop $\tilde t_1$.
Since the gluino contributes to the naturalness measure in Eq.~(\ref{ewnaturalness}) at 2-loop level,
a low fine-tuning allows the gluino with a mass up to several TeV, which is possibly beyond the reach of LHC.
So we assume the gluino mass parameter $M_3=2$ TeV in our scan. Since the electroweak gauginos, heavy Higgs bosons,
the sleptons, the first two generations of squarks and the right-handed sbottom are not strongly related to
the naturalness, we decouple their contributions by fixing $M_1=M_2=m_A=m_{\tilde{\ell}}=m_{\tilde{q}_{1,2}}=m_{\tilde{D}_{R}}=2$ TeV
at weak scale.

In our scan, we impose the following indirect constraints:
\begin{itemize}
\item \textbf{Higgs mass}: We require that the lighter CP-even Higgs boson be the SM-like Higgs boson with a mass in
the range of $125\pm 2$ GeV, which is calculated by the package \textsf{FeynHiggs-2.11.2} \cite{feynhiggs} \footnote{The
prediction of the SM-like Higgs mass depends on the spectrum generators. The differences arise from the choice of the
renormalization scheme and the higher order correction calculations. These effects often lead to a few GeV uncertainty
for the SM-like Higgs mass in the MSSM \cite{higgsmass}.}.

\item \textbf{Vacuum stability}: We impose the constraint of metastability of the vacuum state by requiring $|A_t| \lesssim 2.67\sqrt{M^2_{\tilde{Q}_{3L}}+M^2_{\tilde{t}_R}+M^2_A \cos^2\beta}$ \cite{color-breaking-2}, because the large trilinear parameter $A_t$ can potentially lead to a global vacuum where charge and colour are broken \cite{color-breaking-1,color-breaking-2}.
\item \textbf{Low-energy observables}: We require our samples to satisfy the
bound of $B\rightarrow X_s\gamma$ at 2$\sigma$ range, which is implemented by the package of \textsf{SuperIso v3.3} \cite{superiso}.

\item \textbf{Dark matter detection}: We require the thermal relic density of the neutralino dark matter $\Omega h^2$ is
below the 2$\sigma$ upper limit of 2015 Planck value \cite{planck} \footnote{The thermal relic density of the light
higgsino-like neutralino dark matter is typically low as a result of the large annihilation rate in the early universe.
One possible way to produce the correct relic density is introducing the mixed axion-higgsino dark matter \cite{axion}.
However, if the naturalness requirement is relaxed, the heavy higgsino-like neutralino with a mass $\sim 1-2$ TeV can
solely produce the correct relic density in the MSSM \cite{well-tempered}.} and the LUX WS2014-16 \cite{lux}.
The results for the spin-independent neutralino-proton scattering cross section $\sigma^{SI}_p$ is rescaled by a
factor of $\Omega h^2/\Omega_{PL} h^2$. We use the package of \textsf{MicrOmega v2.4} \cite{micromega} to
calculate $\Omega h^2$ and $\sigma^{SI}_p$.
\end{itemize}

Besides, the LHC run-2 experiments have covered a wide parameter space of the MSSM. We list the relevant LHC
experimental analyses for our scenario:
\begin{itemize}
\item From ATLAS,
\begin{itemize}
\item \textbf{Stop}, 0 lepton + (b)jets + $\slashed E_T$, 13.3 fb$^{-1}$\cite{ATLAS-CONF-2016-077},
\item \textbf{Stop}, 1 lepton + (b)jets + $\slashed E_T$, stop, 13.3 fb$^{-1}$\cite{ATLAS-CONF-2016-050},
\item \textbf{Stop}, 2 leptons + (b)jets + $\slashed E_T$, stop, 13.3 fb$^{-1}$\cite{ATLAS-CONF-2016-076},
\item \textbf{Sbottom}, 2 b-tagged jets + $\slashed E_T$, 3.2 fb$^{-1}$\cite{Aaboud:2016nwl},
\item \textbf{Compressed Spectrum}, 1 jet + $\slashed E_T$, 3.2 fb$^{-1}$\cite{Aaboud:2016tnv}.
\end{itemize}
\item From CMS,
\begin{itemize}
\item \textbf{Inclusive}, 0 lepton + $\geqslant 1$ jets + $\slashed E_T$ + $\alpha_T$, 12.9 fb$^{-1}$\cite{CMS-PAS-SUS-16-016}
\item \textbf{Inclusive}, 0 lepton + $\geqslant 1$ jets + $\slashed E_T$ + $M_{T_2}$, 12.9 fb$^{-1}$\cite{CMS-PAS-SUS-16-015}
\item \textbf{Inclusive}, 0 lepton + $\geqslant 1$ jets + $\slashed E_T$ + $H^{miss}_T$, 12.9 fb$^{-1}$\cite{CMS-PAS-SUS-16-014}
\item \textbf{Stop}, 0 lepton + (b)jets + $\slashed E_T$, 12.9 fb$^{-1}$\cite{CMS-PAS-SUS-16-029},
\item \textbf{Stop}, 1 lepton + (b)jets + $\slashed E_T$, 12.9 fb$^{-1}$\cite{CMS-PAS-SUS-16-028},
\item \textbf{Compressed Spectrum}, 1 jet + soft lepton pair + $\slashed E_T$, 12.9 fb$^{-1}$\cite{CMS-PAS-SUS-16-025}.
\end{itemize}
\end{itemize}
It should be mentioned that the higgsinos $\tilde{\chi}^{\pm}_{1}$ and $\tilde{\chi}^{0}_{2}$ have the small mass difference
with the LSP $\tilde{\chi}^0_{1}$ in NSUSY. Then the decay products of $\tilde{\chi}^{\pm}_{1}$ and $\tilde{\chi}^0_{2}$ are
too soft to be tagged at the LHC. So, the stop decays can be categorized into two topologies: $2b+\slashed E_T$
and $t\bar{t}+\slashed E_T$.
Among the current ATLAS searches for the stop, the all-hadronic final state channel has a better sensitivity than
those with leptons in the high stop mass region ($m_{\tilde{t}_1} > 800$ GeV) because of the application of boosted
top technique. Similar results are obtained by the CMS collaboration. With the decrease of the mass splitting
$\Delta m_{\tilde{t}_1-\tilde{\chi}^0_1}$, the sensitivity of the conventional stop searches for the energetic top
quark in the final states become poor. In particular, if $\Delta m_{\tilde{t}_1-\tilde{\chi}^0_1} \ll m_t$, the stop decay
will be dominated by the four-body channel $\tilde{t}_1 \to bf'\bar{f}\tilde{\chi}^0_1$ \cite{djouadi} or the two-body
loop channel $\tilde{t}_1 \to c \tilde{\chi}^0_1$ \cite{hikasa,margrate,andreas}. Then, the decay products of the stop
are usually very soft so that a high $p_T$ hard jet from the ISR/FSR is needed to tag these compressed stop events,
such as the ATLAS monojet analysis listed above. Note that the very recent CMS monojet with the soft lepton pair
analysis of the compressed electroweakinos can exclude the wino-like chargino mass $m_{\tilde{\chi}^\pm_{1}}$ up to 175 GeV
for a mass difference of 7.5 GeV with respect to the LSP. However, this limit is not applicable to our scenario
because the cross section of the higgsino pair production is 1/4 of the wino pair. On the other hand, both ATLAS
and CMS experiments have performed the inclusive SUSY searches for final states with (generally untagged) jets and
a large amount of $\slashed E_T$, which can also be used to derive limits on the parameter space in various simplified
models. In our study, we reinterpret the recent CMS analysis of $0{\rm -lepton}+(\geqslant 1){\rm jets} + \slashed E_T$.
This strategy is built around the use of the kinematic variable $\alpha_T$, which is constructed from jet-based quantities to provide strong discriminating power between sources of genuine and misreconstructed $\vec{p}^{miss}_T$. Such a variable can highly suppress multijet background, and is suitable for early searches at 13 TeV LHC. Based on the above considerations, we use four LHC experimental analyses to constrain the parameter space of NSUSY, which are listed in Table \ref{table1}.
\begin{table}[h]
\caption{The LHC Run-2 analyses used in our study.}
\begin{tabular}{|c|c|}
\hline~~ ATLAS~~&~~CMS~~\\
\hline
~~1 lepton + (b)jets + $\slashed E_T$ \cite{ATLAS-CONF-2016-050}~~ & ~~0 lepton +($\geqslant 1$)jets + $\slashed E_T$ + $\alpha_T$ \cite{CMS-PAS-SUS-16-016}~~

 \\ \hline
1 jet + $\slashed E_T$ \cite{Aaboud:2016tnv} &  0 lepton + (b)jets + $\slashed E_T$ \cite{CMS-PAS-SUS-16-029}\\
\hline
\end{tabular}\label{table1}
\end{table}

In our Monte Carlo simulations, we use \textsf{MadGraph5\_aMC@NLO} \cite{mad5} to generate the parton level signal events,
which are showered and hadronized by the package \textsf{PYTHIA} \cite{pythia}. The detector simulation effects are
implemented with the package \textsf{Delphes} \cite{delphes}. The jets are clustered with the anti-$k_t$
algorithm \cite{antikt} by the package \textsf{FastJet} \cite{fastjet}. The cross section of the stop pair
production at 13 TeV LHC are calculated by \textsf{NLL-fast} package \cite{nll-fast} with the CTEQ6.6M PDFs \cite{cteq6}.
We impose the ATLAS monojet constraint with \textsf{MadAnalysis 5-1.1.12} \cite{ma5}. The ATLAS 1-lepton stop and
the CMS 0-lepton stop analyses are implemented within the CheckMATE framework \cite{checkmate-1}. But as mentioned above,
we only focus on the heavy stop mass range ($m_{\tilde{t}_1}>500$ GeV) for the CMS 0-lepton analyses because of the improved
sensitivity by application of the top tagging technique.
Besides, the higgsinos $\tilde{\chi}^{\pm}_{1}$ and $\tilde{\chi}^0_{1,2}$ are nearly degenerate in NSUSY. The stop decay
$\tilde{t} \to b \tilde{\chi}^+_1$ gives the same topology as the sbottom decay $\tilde{b} \to b \tilde{\chi}^0_1$.
So we can determine the exclusion limit on the stop by using the cross section upper limit of the sbottom pair production
reported from the CMS inclusive search with $\alpha_T$.

\begin{figure}[h]
\centering
\includegraphics[width=7in]{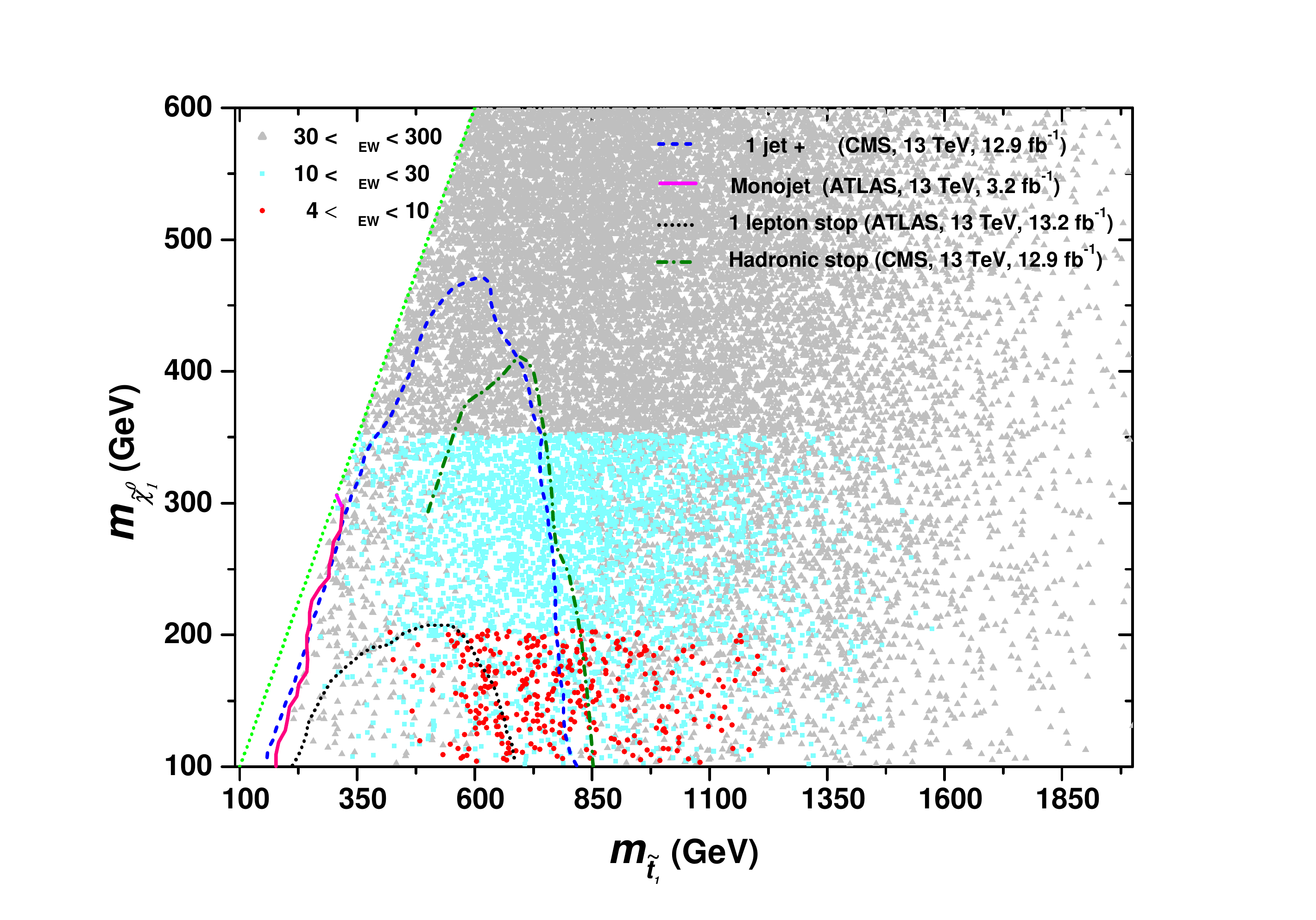}
\vspace*{-1.cm}
\caption{Scatter plots on the plane of $m_{\tilde{t}_1}$ versus $m_{\tilde{\chi}^0_1}$. All samples satisfy the constraints of
the Higgs mass, vacuum stability, $B \to X_s\gamma$ and dark matter detections. The exclusion limits of the LHC SUSY
searches in Table~\ref{table1} are recasted. The triangles (grey), squares (cyan) and bullets (red) represent the
samples that have the electroweak fine-tuning $\Delta_{EW}<10$, $10<\Delta_{EW}<30$ and $30<\Delta_{EW}<300$, respectively.}
\label{result}
\end{figure}
In Fig.~\ref{result}, we project the samples allowed by the Higgs mass, vacuum stability, $B \to X_s\gamma$ and
dark matter detections on the plane of $m_{\tilde{t}_1}$ versus $m_{\tilde{\chi}^0_1}$.
To quantitatively evaluate the naturalness, we use the electroweak fine-tuning measure $\Delta_{EW}$ \footnote{The Barbieri
and Guidice (BG) measure in Ref.~\cite{bg} is applicable to a theory with several independent effective theory parameters.
But for a more fundamental theory, BG measure often leads to an over-estimate of fine-tuning \cite{baer-naturalness-ew}.} \cite{baer-naturalness-ew}
\begin{eqnarray}
\Delta_{EW}\equiv  max_i|C_i|/(M^2_Z/2) ,
\label{ewnaturalness}
\end{eqnarray}
where $C_\mu=-\mu^2$, $C_{H_u}=-m^2_{H_u}\tan^2\beta/(\tan^2\beta-1)$, $C_{H_d}=m^2_{H_d}/(\tan^2\beta-1)$,
$C_{\Sigma_u(i)}=-\Sigma_u(i)(\tan^2\beta)/(\tan^\beta-1)$ and $C_{\Sigma_d(i)}=\Sigma_d(i)/(\tan^\beta-1)$ with $i$ labeling
the various loop contributions to $\Sigma_u$ and $\Sigma_d$. The one-loop stop contributions $\Sigma_u (\tst_{1,2})$
are given by~\cite{peisi}
\begin{eqnarray}
\Sigma_u^u (\tst_{1,2})= \frac{3}{16\pi^2}F(m_{\tst_{1,2}}^2)
\left[ y_t^2-g_Z^2\mp \frac{f_t^2 A_t^2-8g_Z^2
(\frac{1}{4}-\frac{2}{3}x_W)\Delta_t}{m_{\tst_2}^2-m_{\tst_1}^2}\right]
\label{eq:sigtuu}
\end{eqnarray}
where the form factor $F(m^2)= m^2\left(\log\frac{m^2}{Q^2}-1\right)$ with the optimized scale $Q^2 =m_{\tst_1}m_{\tst_2}$,
$y_t$ is the top quark Yukawa coupling and $\Delta_t=(m_{\tst_L}^2-m_{\tst_R}^2)/2+M_Z^2\cos 2\beta (\frac{1}{4}-
\frac{2}{3}x_W)$, $x_W\equiv\sin^2\theta_W$.
In this figure the triangles, squares and bullets represent the samples that have the electroweak fine-tuning
$4<\Delta_{EW}<10$, $10<\Delta_{EW}<30$ and $30<\Delta_{EW}<300$, respectively. In our parameter space, the low
fine-tuning $4<\Delta_{EW}<10$ requires the higgsino mass $\mu \lesssim 200$ GeV and the stop
mass $0.4~{\rm TeV} \lesssim m_{\tilde{t}_1} \lesssim 1.3$ TeV. It can be seen that 70\% of such a parameter space
can be covered by the current LHC Run-2 SUSY searches. A lighter stop mass ($m_{\tilde{t}_1}  \lesssim 0.4~{\rm TeV}$)
requires a large trilinear parameter $A_t$ to satisfy the Higgs mass constraint, which leads to a large value
of $\Delta_{EW}$.

Besides, from Fig.~\ref{result} it can be seen that the ATLAS monojet search produces a strong exclusion limit in
the low stop mass region, which excludes the stop mass up to 320 GeV for $m_{\tilde{\chi}^0_1}=300$ GeV. This is because
that when the stop mass is close to the LSP mass, the $b$-jets from the stop decay $\tilde{t}_1 \to b \tilde{\chi}^+_1/b f\bar{f'}\tilde{\chi}^0_{1,2}$ or $c$-jets from $\tilde{t}_1 \to c \tilde{\chi}^0_{1,2}$ are too soft to be identified.
Then the monojet search is very sensitive in the low stop region.

In the moderate or heavy stop region, the stop dominantly decays to $b \tilde{\chi}^+_1$ and
$t\tilde{\chi}^0_{1,2}$, which produce $2b+E^{miss}_T$ and $t\bar{t}+E^{miss}_T$ signatures, respectively.
The CMS inclusive search with $\alpha_T$ shows a better sensitivity than the 0/1-lepton stop searches in most
parameter space. But we also note that the exclusion limit of the CMS 0-lepton stop search is slightly stronger
than the CMS inclusive search because of the application of top tagging technique in ATLAS analysis. Finally,
we conclude that the stop mass can be excluded up to 765 (850) GeV for $m_{\tilde{\chi}^0_1} < 300$ ($m_{\tilde{\chi}^0_1} = 100$)
GeV by the current LHC Run-2 experiments. Such limits are much stronger than the LHC run-1 limits on NSUSY,
which excluded a stop below 600 GeV \cite{wu-2,kim-1,wu-1,warsaw}.

It should be mention that when the stop and LSP mass splitting $\Delta m_{\tilde{t}_1-\tilde{\chi}^0_1} \simeq m_t$, the kinematics of the top quarks from stop decay are similar to those in the top pair production so that the above LHC searches for stop pair have the poor sensitivity. With the help of an additional high momentum jet recoiling against stop pair system, one can utilize the observable $R_M\equiv \slashed E_T/p_T(j_{ISR})$ to extend the reach of stop to about 800 GeV at 13 TeV LHC with ${\cal L}=3000$ fb$^{-1}$ \cite{liantao}. Besides, the VBF production of the stop pair was also proposed to detect such a compressed stop region, which can cover the stop mass to about 300 GeV because of the large systematical uncertainty \cite{dutta}. In NSUSY, when both decay channels $\tilde{t}_1 \to t \tilde{\chi}^0_1$ and $\tilde{t}_1 \to b \tilde{\chi}^+_1$ are allowed, search for the asymmetric final states $\tilde{t}(\to t \tilde{\chi}^0_1)\tilde{t}^\dagger(\to b\tilde{\chi}^-_1)$ can provide a complementary way to probing stop at the LHC. With the variable topness to suppress $t\bar{t}$ background, such asymmetric stop search has a comparable sensitivity with the symmetric stop searches at the HL-LHC \cite{topness}. Therefore, together with conventional LHC search strategies, we can expect that the future high luminosity LHC is able to probe the stop and higgsino mass up to 1.5 TeV and 0.6 TeV, respectively \cite{kim-2}. At that time, most of the NSUSY parameter space with $\Delta_{EW}<30$ can be covered \cite{kim-2,Baer-HL}.

\section{conclusions}
In this paper, we examined the lower mass limit of the stop in natural supersymmetry (NSUSY) by using the recent
LHC-13 TeV data. We recast the LHC SUSY inclusive search for $(\ge 1){\rm jets}+ \slashed E_T$ events with $\alpha_T$
variable, the direct stop pair searches (1-lepton channel and all-hadronic channel) and the monojet analyses. We found
that the inclusive SUSY analysis with $\alpha_T$ is complementary to the direct stop pair analyses in probing NSUSY.
The current LHC data can exclude the stop up to 765 (860) GeV for $m_{\tilde{\chi}^0_1} < 300$ ($m_{\tilde{\chi}^0_1} = 100$) GeV. While in the compressed region ($\Delta m_{\tilde{t}_1-\tilde{\chi}^0_1} \simeq 20$ GeV), the stop mass can be still light as
320 GeV. About 70\% of the NSUSY parameter space with $\Delta_{EW}<10$ can be covered by the current LHC Run-2 data. The future HL-LHC is expected to push the lower mass limits of the stop and higgsino up to 1.5 TeV and 0.6 TeV, respectively and cover most NSUSY parameter space with $\Delta_{EW}<30$.

\acknowledgments
This work is partly supported by the Australian Research Council,
by the CAS Center for Excellence in Particle Physics (CCEPP),
by the National Natural Science Foundation of
China (NNSFC) under grants Nos. 11275057, 11305049, 11375001, 11405047, 11135003, 11275245,
by Specialised Research Fund for the Doctoral Program of Higher Education under Grant No. 20134104120002.
CCH is supported by World Premier International Research Center Initiative (WPI Initiative), MEXT, Japan.
MCZ is supported by Institute for Basic Science (IBS-R018-D1).

\end{document}